\documentclass[aps,prd,amsfonts,nofootinbib,supercriptaddress,11pt]{revtex4}
\usepackage{amssymb,amsmath}

\usepackage{graphicx}
\usepackage{epstopdf}

\usepackage{epsfig}
\usepackage{dcolumn}   
\usepackage{bm}        
\usepackage{amssymb}   

\newcounter{fig}

\newcommand{\beq}{\begin{equation}}
\newcommand{\eeq}{\end{equation}}
\newcommand{\bea}{\begin{eqnarray}}
\newcommand{\eea}{\end{eqnarray}}

\usepackage{amsmath}

\begin{document}



\title{Global monopoles can change Universe's topology}

\author{Anja Marunovi\'c}
\author{Tomislav
Prokopec\footnote{a.marunovic@uu.nl, t.prokopec@uu.nl}}

\affiliation{Institute for Theoretical Physics, Spinoza Institute
and $EMME\Phi$,
 Utrecht University,\\
Leuvenlaan 4, 3584 CE Utrecht, The Netherlands}

\begin{abstract} \noindent

 If the Universe undergoes a phase transition, at which
global monopoles are created or destroyed, topology of its
spatial sections can change.
More specifically,
by making use of Myers' theorem,
we show that, after a transition in which global monopoles form,
spatial sections of a spatially flat, infinite Universe become finite and closed.
This implies that global monopoles can change
the topology of Universe's spatial sections
(from infinite and open to finite and closed).
Global monopoles cannot alter the topology
of the space-time manifold.

\end{abstract}


\maketitle

\vskip -0.5cm
\section{Introduction}
\label{Introduction} \vskip -0.1cm

 The question of global properties (topology) of our Universe is a fascinating one, and
it has been attracting attention for a long time. 
Yet only as-of-recently the data have been good enough to put meaningful observational constraints on
the Universe's topology. While Einstein's equations uniquely specify local properties of space-time 
(characterised by the metric tensor), 
they fail to determine its global (topological) properties.  
Friedmann, Robertson and  Walker (FRW) were first who observed that
the most general solution corresponding to spatially homogeneous  Universe with constant 
curvature $\kappa$ of its spatial sections is the following FLRW metric ($L$ stands for Lemaitre),
\begin{equation}
ds^2 = -c^2dt^2  + \frac{a^2(t)dr^2}{1-\kappa r^2}
   +a^2(t) r^2 [d\theta^2+ \sin^2(\theta)d\phi^2],
\label{general FLRW solution}
\end{equation}
where $c$ is the speed of light and $0\leq r<\infty, 0\leq \theta\leq \pi,0\leq \phi<2\pi$
are spherical coordinates. Recent cosmic microwave background and large scale structure observations 
tell us that, at large scales the metric~(\ref{general FLRW solution})
describes quite accurately our Universe. When $\kappa$ in~(\ref{general FLRW solution}) is
\begin{enumerate}
\item  negative ($\kappa <0$), then
 spatial sections of the Universe are hyperbolic,
\item zero ($\kappa =  0$), then spatial sections are flat;
\item positive ($\kappa >  0$), then the spatial sections are positively curved and they are locally homeomorphic to the geometry of the three dimensional sphere.
\end{enumerate}

 Older literature typically assumes that  $\kappa \leq 0$ implies infinite spatial sections, while 
when $\kappa > 0$, spatial sections are compact. While the latter statement is correct, recent advancements in 
our understanding of (topology of) three dimensional manifolds tell us that we must be much more 
careful when drawing conclusions from the observational fact that the metric describing our observable Universe is well approximated 
by the FLRW metric~(\ref{general FLRW solution}).
Namely, various boundary conditions could be imposed on the Universe's spatial sections~\cite{Levin:2001fg}, 
giving as a result a large number of possible three dimensional manifolds, only one of which corresponds to that of our Universe.

Let us now briefly recall the relevant observational facts.
 The first observational evidence that supports that we live
in a (nearly) flat universe ($\kappa\approx 0$) was presented in 2000 by the balloon
experiments Boomerang~\cite{de Bernardis:2000gy} and
Maxima~\cite{Balbi:2000tg}.
A recent bound on $\kappa$~\cite{Aubourg:2014yra}
is obtained when observations of BAOs (Baryon Acoustic Oscillations) are combined with
the Planck data~\cite{Ade:2013zuv} and the
polarization data from the WMAP satellite (WP),
\vskip -0.7cm
\beq
 \Omega_\kappa = -0.003\pm 0.003,
\,,\qquad \Omega_\kappa = -\frac{\kappa c^2}{H_0^2}
\,,\label{bound on kappa: 2014}
 \eeq
\vskip -0.28cm\noindent
where $H_0\simeq 68~{\rm km/s/Mpc}$
(when the BAO data are dropped,
one obtains $0.006>\Omega_\kappa>- 0.086$~\cite{Ade:2013zuv}).
Eq.~(\ref{bound on kappa: 2014}) implies a large lower bound on the
curvature radius of spatial sections,
$R_{\rm c}=1/\sqrt{|\kappa|}\geq 60~{\rm Gpc}$.
The bound~(\ref{bound on kappa: 2014}) implies the following  
robust conclusion: "our [observable] Universe is
spatially flat to an accuracy of better than a percent" (cited from
page 42 of Ref.~\cite{Ade:2013zuv}).

Even if the Universe is spatially flat, it can be made finite 
by imposing suitable periodic boundary conditions; the precise nature of periodic conditions 
determins Universe's global topology~\cite{Levin:2001fg}. Although different scenarios
have been considered in literature (good reviews
are given in Refs.~\cite{Ade:2013vbw,Ade:2015bva,Levin:2001fg,Uzan:1998hk,Luminet:1999qh,Starkman:1998qx}),
so far no evidence has been found
that would favor any of the proposed models. For example,
extensive mining of the CMB data have been
performed~\cite{Vaudrevange:2012da,Cornish:2003db,Cornish:1997ab}
in order to find pairs of circles, which are a telltale signature for non-trivial large-scale topology 
of the Universe, but so far no convincing signature has been found.

 The above considerations make an implicit assumption that spatial curvature
of the Universe is given and that it cannot be changed throughout the history of
our Universe. In this letter we argue that this assumption ought to be relaxed,
and we propose a dynamical mechanism:
\begin{enumerate}
\item[] {\it formation of global monopoles at an early universe phase transition,}
\end{enumerate}
by which the (average, measured) spatial
curvature of the Universe can change in the sense that it will become positive
if it starts slightly negative or zero. Strictly speaking this is true
provided the Universe was before the transition non-compact, {\it i.e.}
it was created with no periodic boundary conditions imposed on it.

 This claim will leave many readers with a queasy feeling since, when
$\kappa$ changes from $\kappa\leq 0 $ to $\kappa>0$, spatial
sections could change from infinite (hyperbolic or parabolic) to finite
(elliptic), thus changing the topology of spatial sections.  One should keep in mind
that all this happens at \emph{space-like} hyper-surfaces of
constant time, and hence it is not in contradiction with any laws of
causality. And yet it does leave us with an uncomfortable feeling
that `somewhere there' distant spatial sections of the Universe are
{\it reconnecting}, thereby changing them from infinite to finite
and periodic. This will be the case provided all spatial dimensions are equally affected, which is the case
in the mechanism considered in this letter.
Even though not directly observable today, this
process can have direct consequences for our future. Indeed, when
an observer in that reconnected Universe sends a (light) signal, it
will eventually arrive from the opposite direction. Furthermore, the
future of a spatially finite (compact) universe can change from
infinite and uneventful to finite and singular (namely,  if
cosmological constant is zero such a universe will end up in a Big
Crunch singularity). Because of all of these reasons, a tacit
consensus has emerged that no topology change is possible in our
Universe (albeit strictly speaking measurements constrain the Universe's spatial topology only after recombination).
We argue in this work that this consensus needs to be
reassessed.

In fact, the idea that the curvature of spatial sections could
change can be traced back to the work of
Krasinski~\cite{Krasinski:1982zz} based on Stephani's exact
solution~\cite{Stephani:1967} to Einstein's equations. Even though
Krasinski has argued that the curvature of spatial sections could
dynamically change, he has not offered any mechanism by which such a
change could occur~\cite{footnote1}. In this letter we provide such
a dynamical mechanism.

 A particularly instructive case to consider is the maximally symmetric de Sitter space,
whose geometry can be clearly visualized from its five dimensional
(flat, Minkowskian) embedding (see figure 1),
\bea
  &&dS^2  = -dT^2 + dX_1^2 + dX_2^2 + dX_3^2 + dX_4^2
\,,\nonumber \\ && R_H^2 = T^2 - R^2 \,,\,\,  R^2 = X_1^2 + X_2^2 +
X_3^2 + X_4^2 \,. \label{dS:embedding} \eea
Thus de Sitter space is geometrically a four dimensional hyperboloid
$\mathbb{H}^4$,
and its symmetry is the five dimensional Lorentz group, $SO(1,4)$, which
has -- just like the Poincar\'e group of the symmetries of Minkowski space --
10 symmetry generators.
This means  that de Sitter space also has 10 global symmetries
(Killing vectors).  Common coordinates on de Sitter
space~(\ref{dS:embedding}) are those of constant
curvature of its spatial sections, and they include:
(a) closed (global) coordinates have $\kappa>0$;
(b) flat (Euclidean) coordinates (Poincar\'e patch) have $\kappa=0$ and
(c) open coordinates (hyperbolic sections) have $\kappa<0$.
 Krasinski has, however, pointed out that
there are also de Sitter coordinates in which $\kappa$ changes in time.
Both cases, when $\kappa$ changes from negative to positive, and
{\it v.v.} are possible. An example of the metric when $\kappa(t)$
changes from negative to positive can be easily inferred
from~\cite{Krasinski:1982zz},
\vskip -0.5cm
\bea
  ds^2 & =&
  -\frac{c^2(r/r_0)^4}{[1+ctr^2/r_0^3]^2[(Hr_0/c)^2-ct/r_0]}dt^2\nonumber
  \\
    &&+\frac{1}{[1+ctr^2/r_0^3]^2}
    \Big[dr^2 + r^2  d\theta^2+ \sin^2(\theta)d\phi^2\Big]\,,
\quad
     \label{dS:kappa time dependent}
 \eea
\vskip -0.15cm\noindent
where $r_0, c, H$ are constants.
That this is a de Sitter space can be checked, for example, by evaluating
the Riemann tensor. One finds
\vskip -0.5cm
\begin{equation}
R_{\mu\nu\alpha\beta}=({R}/{12})(g_{\mu\alpha}g_{\nu\beta}
     -g_{\mu\beta}g_{\nu\alpha})\,,
\label{Riemann tensor:dS}
\end{equation}
\vskip -0.06cm\noindent
where $R= 12H^2/c^2\equiv 12/R_H^2$ is the Ricci curvature scalar,
 $H={\rm const.}$ is the Hubble parameter and $R_H=c/H$ is the Hubble radius.
Relation~(\ref{Riemann tensor:dS}) holds uniquely
for maximally symmetric spaces such as de Sitter space.
The curvature of spatial sections of de Sitter in~(\ref{dS:kappa time dependent})
can be inferred from the Riemann curvature of spatial sections,
\vskip -0.6cm
\begin{equation}
\!^{(3)}\!R_{ijkl}=\frac{ ^{(3)}\!R}{6}(g_{ik}g_{jl}\!-\!g_{il}g_{jk}) \,,\,\,
^{(3)}\!R = \frac{24ct}{r_0^3}\equiv \frac{6\kappa(t)}{a^2(t)},\!
\label{Riemann tensor:dS2}
\end{equation}
\vskip -0.3cm\noindent from which we infer, \vskip -0.5cm
\begin{equation}
      \kappa(t) = \frac{4cta^2(t)}{r_0^2}
\,,\quad \mbox{with}\quad a(t) = {\rm e}^{Ht} \,,
\end{equation}
\vskip -0.16cm \noindent which means that $\kappa<0$ for $t<0$,
$\kappa=0$ for $t=0$, and $\kappa>0$ for $t>0$. Note that topology
of spatial sections changes at $t=0$. For $t<0$ the sections are
three dimensional hyperboloids, with a time dependent (physical)
throat radius $r_c(t)=r_0^{3/2}/\sqrt{-ct}$, for $t=0$ they are
paraboloids and for $t>0$ they are three-spheres with a
(time-dependent) radius, $r_c=r_0^{3/2}/\sqrt{ct}$ (see figure~1).
Consequently, topology of spatial sections changes at $t=0$, as can
be seen in figure~1~\cite{footnote2}. A similar (albeit
inhomogeneous) construction is possible on FLRW space-times. While
this shows that there are observers for which topology of spatial
sections of an expanding space-time changes, it does not tell us how
to realize such a change, and whether such a change is possible in a
realistic setting. This is what we address next.

\begin{figure}
\centering
\includegraphics[scale=0.5]{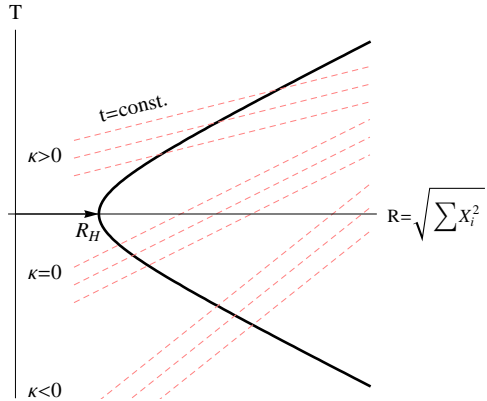}
\caption{Hypersurfaces of constant time of de Sitter $\mathbb{H}^4$ with a time
dependent $\kappa$.} \vskip -0.3cm
 \label{fig}
\end{figure}
\vskip -0.5cm

\section{The Model}
\label{The Model} \vskip -0.1cm

The action for gravity we take to be the Einstein-Hilbert (EH)
action,
\vskip -0.6cm
\beq S_{EH}=\frac{c^4}{16\pi G_N}\int d^4 x \sqrt{-g}R \,,
\label{E-H action} \eeq
\vskip -0.16cm\noindent
where $G_N$ is the Newton constant, $R$ is the Ricci curvature
scalar, and $g$ is the determinant of the metric tensor
$g_{\mu\nu}$. In our model the EH action is supplemented by the
action that governs the dynamics of global monopoles,
\vskip -0.6cm
\beq
S_\phi=\int d^4 x \sqrt{-g}\left(
            - \frac 12 g^{\mu\nu}(\partial_\mu\phi^a)(\partial_\nu\phi^a)
            - V(\phi^a)
                       \right),
\label{Monopole Action}
\eeq
\vskip -0.16cm\noindent
with the Higgs type of $O(3)$ symmetric potential
\vskip -0.6cm
\beq
V(\phi^a)=\frac{\mu^2}{2}\,\phi^a\phi^a
   +\frac{\lambda_\phi}{4}(\phi^a\phi^a)^2+\frac{\mu^4}{4\lambda_\phi}
\,,
\label{potentialA}
\eeq
\vskip -0.16cm\noindent
where repeated indices $a$ indicate a summation over $a=1,2,3$.
$\mu$ is a mass parameter and $\lambda_\phi$ is a self-coupling. The
scalar field $\vec \phi=(\phi^a)$ ($a=1,2,3$) consists of 3 real
components, such that the action~(\ref{Monopole Action}) is
$O(3)$-symmetric. When $\mu^2<0$ the vacuum exhibits a field
condensate, $\phi^a\phi^a\equiv \phi_0^2=-\mu^2/\lambda_\phi$, which
spontaneously breaks the $O(3)$ symmetry of the action to an $O(2)$,
such that the resulting vacuum manifold ${\cal M}$ has a symmetry of
the two dimensional sphere, ${\cal M}=O(3)/O(2)\sim S^2$. The two
excitations along the two orthogonal directions of $S^2$ are the two
massless Goldstone bosons, while the excitation orthogonal to ${\cal
M}$ is massive, $m^2(\phi_0)= 2\lambda_\phi\phi_0^2=-2\mu^2$. The
potential~(\ref{potentialA}) is chosen such that the energy density
of the (classical) vacuum is $V(\phi_0)=0$.

It is well known that the action~(\ref{Monopole
Action}--\ref{potentialA}) permits non-vacuum classical solutions
known as global monopoles~\cite{Kibble:1976sj}. Global monopoles
have a non-vanishing vacuum energy, but they do not decay as they
are stabilised by topology. The simplest such solution of the
equations of motion is a hedgehog-like spherically symmetric
solution of the form,
\vskip -0.6cm
\beq
\vec \phi(t,\vec r\,)
  = \phi(r)\left(\sin\theta\,\cos\varphi,\sin\theta\,\sin\varphi,
      \cos\theta\right)^T
\,,
\label{AnsatzSca}
\eeq
\vskip -0.16cm\noindent
where $\theta, \varphi$ and $r$ are
spherical coordinates. One can show~\cite{Marunovic:2014}
 that the topological charge (also known as the winding number)
of that solution is unity,
\vskip -0.5cm
\begin{equation}
  Q[\phi^a] = \frac{1}{8\pi}\int dS^{ij}
\frac{\epsilon^{abc}\phi^a\partial_i\phi^b\partial_j\phi^c}{[\phi^e\phi^e]^{3/2}}
        = 1,\,\,
 dS^{ij}= dx^i \wedge dx^j
\,,
\label{topological charge = 1}
\end{equation}
\vskip -0.16cm\noindent
and that it is stable under small field perturbations.

 Global monopoles are generically created at a phase transition
 by the Kibble mechanism~\cite{Kibble:1976sj}
(at least of the order one per Hubble volume) if the effective field
mass matrix, $[m^{2}_{\rm eff}(\phi^a=0)]^{ab} =\partial^2V_{\rm
eff}/\partial\phi^a\partial\phi^b$ changes from having all positive
eigenvalues to at least one negative
eigenvalue~\cite{Prokopec:2011ms,Lazzari:2013boa}, which can be
realised in {\it e.g.} a hot Big Bang. Global monopoles with charge
$Q=1$ and $Q=-1$ (which can be obtained by interchanging any two
coordinates in~Eq.~(\ref{AnsatzSca})) are equally likely to form,
and the monopoles of opposite charge will strongly attract each
other (by a force independent of distance) and efficiently
annihilate~\cite{Leese:1990cj}, such that a network of global
monopoles will eventually reach a scaling
solution~\cite{Bennett:1990xy} with about four monopoles per Hubble
volume at any given time.

One important and defining property of global monopoles is their
{\it solid deficit angle}~\cite{Barriola:1989hx}
(see also the Appendix), which extends to the particle
horizon associated with the monopole creation event.


\vskip -0.2cm\noindent

\section{Global and local properties of space-time}
\label{Global and local properties of space-time} \vskip -0.1cm

As shown in the Appendix, at sufficiently large distances, the
metric of a monopole can be approximated by a FLRW metric with a
solid deficit angle. This deficit angle generates a spatial Ricci
tensor that breaks spatial homogeneity of the FLRW space-time and
decays as $ ^{(3)}\!R_r^r\simeq 4\Delta/[m_\phi^2a^4r^4]$,
 $ ^{(3)}\!R_\theta^\theta=^{(3)}\!\!R_\varphi^\varphi
   \simeq \Delta/[a^2(t)r^2]$,
where $\Delta$ is the deficit angle and $m_\phi$ the
monopole mass.
In the limit of a large number of monopoles
per Hubble volume all with $\Delta\ll 1$ and if in the absence of
monopoles the spatial curvature is zero, a local observer that is
sufficiently far from any individual monopole will observe a metric
that can be approximated by a FLRW metric with a positive spatial
curvature. Hence, an observer in such a universe will tend to
conclude that the Universe is spatially compact and finite.

In this section we show that a (spatially flat) cosmological space
filled with randomly distributed global monopoles  must have spatial
sections that are closed, and therefore its spatial geometry is that
of a three dimensional sphere. This conclusion is reached based on
the well known \emph{Myers' theorem}~\cite{Myers:1941}, which states
that for any Riemannian manifold whose Ricci curvature $R_i^{\,j}$
is positive and limited from below as,
\vskip -0.6cm
\begin{equation}
 \|R_i^{\,j}\|> R_{\rm min}>0
\label{Myers thm}
\end{equation}
\vskip -0.16cm\noindent
the distance function $d(x;x')$ is limited from above by
$d(x;x')<\pi/\sqrt{R_{\rm min}}$
 in any number of (spatial) dimensions.
This then implies that the manifold is {\it globally closed} and its radius of
 curvature is limited from above as $r_c \leq\pi/\sqrt{R_{\rm min}}$.
This powerful theorem relates local properties of Riemannian
manifolds to their global properties. In particular, when applied to
global monopoles,
that, at asymptotically large distances, the minimum component
of the Ricci is  (see Eq.~(\ref{Ricci scalar: asymptotic})
in the Appendix), \vskip -0.6cm
\begin{equation}
 \|^{3}\!R_i^{\,j}\|=\!\! \phantom{,}^{(3)}R_r^r
      \simeq \frac{4\Delta}{m_\phi^2a^4r^4}
 \,.
\label{Ricci:upper bound}
\end{equation}
\vskip -0.3cm\noindent
Assume that the monopole-monopole
correlation function is that of randomly (Poisson)
distributed monopoles with an average distance squared, $\langle
a^2(\vec x-\vec x')^2\rangle  \simeq a^2\sigma^2$. Let us consider
a sphere of radius $ar\sim a \sigma N^{1/3}$, in which the average number
of monopoles is, $N\sim (ar/\sigma)^3\sim 1/\Delta$, which is the
number needed to close the space (here we neglect factors of order
unity, such as the volume of the 3-dimensional unit sphere, ${\cal
V}ol(S^3)=2\pi^2$). The Poisson distribution then implies that the
maximum distance between two monopoles is
$ar_{\rm max}\sim a\sigma [\ln(\Delta^{-1})]^{1/3}$. Now, according
to Myers' theorem, a spatial section of the Universe filled with
monopoles is a closed inhomogeneous manifold with a (comoving) radius of
curvature, $r_c\lesssim 
m_\phi a\sigma^2 \Delta^{-1/2}[\ln(\Delta^{-1})]^{2/3}$. While this
represents an upper bound, the actual  Universe's curvature radius
will be smaller, probably of the order $r_c\sim
\sigma/\sqrt{\Delta}$, where a dilute monopole gas is assumed, $m
a\sigma\gg 1$.

To conclude, due to their deficit angle, formation of global monopoles can
have an impact on global (topological) properties of the
(spatial sections of the) Universe.
In particular, a (spatially flat) universe filled with global
monopoles will have a closed geometry and, in the limit of many monopoles where
each has a small deficit angle,
it will closely resemble a closed universe with a constant positive spatial
curvature $\kappa$. We have thus shown that, if before a phase transition
at which global monopoles form, spatial sections of the Universe are flat
or slightly negatively curved and therefore can be (in the absence of non-trivial topology) infinite,
after the phase transition the Universe will have on average a positive curvature
and its spatial sections will be homeomorphic to a three-sphere
and hence compact.

Even though the Universe filled with global monopoles resembles a
FLRW universe with $\kappa>0$, it is an inhomogeneous universe
with an uncertain future (at the moment
it is not clear to us whether the Universe will end up in a Big
Crunch or it will expand forever). Furthermore, an observer residing
sufficiently close to a monopole, on top of the usual Hubble flow
will feel a repulsive gravitational force that points away from the
monopole core (see e.g.\cite{Marunovic:2013eka}). By studying physical
effects of this force a local observer will be able to distinguish
between a homogeneous universe and a universe filled with global
monopoles.
Next,  while on a clump of matter photons and particles get deflected
by an angle that depends on their velocity,
sufficiently far from a global monopole the deflection angle will be equal for
(relativistic and nonrelativistic) massive particles and photons, {\it i.e.}
it will be independent on particle's speed.

Furthermore, measuring spatial curvature on the largest observable scales
(such as it is done by modern CMB observations) can provide
an upper limit on the number of monopoles in our horizon.
Since CMB observatories such as the Planck
and WMAP satellites measure $\kappa$ on the Hubble scale, they can be
reinterpreted as the upper limit on
the total solid deficit angle in our Hubble volume generated by global monopoles,
\vskip -0.6cm
\begin{equation}
  0<\!\!\!\!\sum_{i\in\rm Hubble\; shell\;
thickness\;\sigma}\!\!\!\!\Delta_i < -(\Omega_\kappa)_{\rm max}\sim
0.01 \,, \label{global monopoles: limit}
\end{equation}
\vskip -0.25cm\noindent
where $-(\Omega_\kappa)_{\rm max}$ represents the upper limit on $-\Omega_\kappa=\kappa c^2/H_0^2$
allowed by the observations and $H_0\simeq 68~{\rm km/s/Mpc}$
is the Hubble parameter today.

\section{Discussion}
\label{Discussion}

 In this letter we show that formation of global monopoles in the early Universe can lead to a change in
(average) spatial curvature $\kappa$ of the Universe such that, if the Universe starts with
 an average $\kappa=0$, after the phase transition, $\kappa>0$.

 An important question is whether our model is consistent with all of the current observations.
As mentioned above, global monopoles eventually reach a scaling solution with
about 4 monopoles per Hubble volume. Each global monopole will generate a perturbation
in matter density that corresponds to a gravitational potential of the order $\psi_N\sim \Delta$,
implying that  $\Delta$ must be smaller than about $10^{-5}$ (which is the amplitude of
a typical potential generated by inflationary perturbations).
A mild breakdown of spatial homogeneity and isotropy is consistent
with CMB observations, and may be related with some of the
CMB anomalies, examples being the observed tantalizing hints for
non-Gaussianities~\cite{Ade:2013nlj,Ade:2015lrj}.

Furthermore, above we found that a  typical spatial curvature
generated by monopoles is of the order, $\kappa\sim
\Delta/\sigma^2\sim (N_M^{2/3}\Delta)(aH)^2$, which increases
(decreases) in accelerating (decelerating) space-times, where $N_M$
is the number of monopoles in the observational volume ($N_M\sim 4$
if the observational volume equals the Hubble volume). This then
implies that the spatial curvature induced by global monopoles in
our Hubble volume is of the order, $\kappa \sim \Delta$, which is
about 2 orders of magnitude too small to be observable. A way out is
to have $N$ copies of scalar fields, each with an $O(3)$ symmetric
potential and leading to formation of global monopoles. In this case
each of the corresponding global monopole networks would be today in
the scaling solution, implying that $\kappa\sim (4N)^{2/3}\Delta$
per Hubble volume. Thus one could get $\kappa\sim 10^{-3}$ (which is
observable) with $N\sim 250$ copies, where we took $\Delta\sim
10^{-5}$. Such a large number of scalar fields can occur {\it i.e.}
in some string compactification models.

 In this letter we focus mainly on the case when the very early Universe 
is non-compact and when its spatial sections have 
a constant negative or vanishing spatial curvature and argue that 
a phase transition at which global monopoles form can change
the perceived spatial curvature to positive and that the spatial 
sections may thus become compact. We are very much aware of 
that this model of the Universe may be too simplistic and 
that the true geometry of the spatial sections of the early Universe 
may be much more complex. Firstly, the Universe's spatial sections 
could be inhomogeneous and could be made up of piecewise 
connected simple three dimensional 
Thurston geometries~\cite{Thurston:1982zz} 
that are used to classify three dimensional manifolds.
The eight Thurston's classes of geometries are: (1) hyperbolic ($H^3$),
(2) Euclidean ($E^3$), (3) spherical ($S^3$), (4) $H^2\times R$, (5)
$S^2\times R$, (6) $SU(2,R)$, (7) NIL geometry, and (8) Solv geometry.
This letter is not the place to address in depth the properties of any of
these geometries (see Ref.~\cite{Levin:2001fg} for an incomplete account of these geometries; 
a more complete, but more mathematical, discussion can be found in 
Refs.~\cite{Weeks:1985,Thurston:1982zz}).
One should note, however, that spatial slices can correspond to 
the covering spaces of these geometries (in which case no periodic boundary 
conditions are imposed) or periodic conditions are imposed on them, which would make 
the Universe's spatial sections compact, and complicate the discussion of what happens 
when global monopoles form. This letter focuses primarily on the simple case 
in which a non-compact $H^3$ spatial geometry changes via $E^3$ to $S^3$ 
(assuming no periodic boundary conditions). 
If however the space is hyperbolic and compact 
(that can be achieved by imposing suitable periodic conditions on a compact fundamental domain)
formation of topological defect will in general {\it not} change the topology of the Universe. 
This is so because compact $H^3$ or $H^2$ topologies have as the fundamental domain a compact space that 
cannot be smoothly changed to that of a flat or positively curved fundamental domain.
To illustrate the point, consider $H^2\times R$. The fundamental  domains that correspond to 
a compact subspace $H^2$ comprise  
polygons $P_{4g}$, where $g\geq 2$ denotes the genus of the manifold. 
It is now clear that these domains cannot be smoothly changed to $P_4$ or $P_6$, which 
correspond to the fundamental domains of flat two dimensional spaces.
Finally, in the 
case when the topology of the Universe belong to one of the last 3 Thurston classes 
($SL(2,R)$, NIL or Solv), since they do not have the positively curved counterparts, we 
expect that a local observer will not be able to infer a change of topology from a non-compact one to 
a compact one. It is clear that further study is needed to fully understand 
the dynamics of topology of the spatial sections of our Universe.

\section*{Appendix}
\vskip -0.3cm \label{Appendix}

The energy-momentum tensor of a global monopole is given by
\vskip -0.6cm
 \bea
T_{\mu\nu}^\phi = -\frac {2}{\sqrt{-g}}\frac{\delta S_\phi}{\delta
g^{\mu\nu}}
  =  (\partial_\mu \phi^a)(\partial_\nu\phi^a)
 \label{tt}
+g_{\mu\nu} {\cal L}_\phi\,.\;
%
 \eea
\vskip -0.2cm\noindent Let us assume an $AB$ metric of the form,
\vskip -0.5cm
\beq
ds^2=-c^2dt^2\!+\!A^2(t,r)dr^2\!+\!B^2(t,r)r^2 [d\theta^2
\!+\! \sin^2(\theta)d\phi^2]
\,.
\label{AB metric}
\eeq
\vskip -0.02cm\noindent In presence of a global
monopole~(\ref{AnsatzSca}) the asymptotic behaviour of the $A$ and
$B$ functions in~(\ref{AB metric}) is~\cite{Marunovic:2014}, \vskip
-0.6cm
\begin{equation}
 \!\,A^2\stackrel{r\rightarrow \infty}{\longrightarrow}
\frac{a^2(t)[1\!+\!f_0/r^2]}{1\!-\!\Delta}
,\;  B^2\stackrel{r\rightarrow
\infty}{\longrightarrow} a^2(t)\bigg(\!1\!+\!\frac{g_0}{r^2}\!\bigg)\!\!\!
\label{AB:asymptotics}
\end{equation}
\vskip -0.2cm\noindent ($f_0$ and $g_0$ are constants) and that of
$\phi(r)$ in~(\ref{AnsatzSca}) is, \vskip -0.6cm
\begin{equation}
\!\,\phi(0)=0,\; \phi(r\rightarrow
\infty)=\phi_0-\frac{1}{\lambda_\phi\phi_0a^2r^2},
\label{phi:asymptotics}
\end{equation}
\vskip -0.2cm\noindent with $\phi_0^2=-\mu^2/\lambda_\phi$, see
Eqs.~(\ref{Monopole Action}--\ref{AnsatzSca}). With these in mind,
one obtains, for non-vanishing components of the stress-energy
tensor the following asymptotic form, \vskip -0.6cm
\bea \! \, && T_r^r  \! \stackrel{r\rightarrow
\infty}{\longrightarrow}\!
    - \frac{\phi_0^2(1\!-\!2/m_\phi^2a^2r^2)^2}{a^2r^2(1+g_0/r^2)} \!-\! \frac{2\phi_0^2}{m_\phi^2a^4r^4}
,\;\,\nonumber\\
&& T_\theta^\theta \!=\!T_\varphi^\varphi\stackrel{r\rightarrow
\infty}{\longrightarrow}\!- \frac{2\phi_0^2}{m_\phi^2a^4r^4},\!\!
 \label{Tmn:asymptotic form}
\eea
\vskip -0.3cm\noindent where $m_\phi^2 =2 \lambda_\phi\phi_0^2\gg
H^2$. On the other hand, the Einstein  equations for the manifold's
spatial sections are,
 \vskip -0.5cm
\begin{equation}
 \!\! \phantom{,}^{(3)}R_{i}^{\;j}
  - \frac{\phantom{\!}^{(3)}R}{2}\delta_{i}^{\;j}\simeq\!
    \frac{8\pi G_N}{c^4}T_i^{\;j}
\Rightarrow
  \phantom{\!}^{(3)}R\!  =  -\frac{16\pi G_N}{c^4}\sum_iT_i^{\;i}
\label{Einstein equation: spatial components}
\end{equation}
\vskip -0.03cm\noindent Upon comparing these
with~(\ref{Tmn:asymptotic form}) one obtains that the monopole
contributes to the relevant components of the spatial Ricci
curvature tensor as, \vskip -0.5cm
\begin{equation}
\!\!\!\!\!\!
  \phantom{X}^{(3)}\!R_r^{\,r} \!\simeq\!   \frac{4\Delta}{m_\phi^2a^4r^4}
,\!\!
  \phantom{X}^{(3)}\!\!R_\theta^{\,\theta} \!=\! \!\! \!\!\!
\phantom{X}^{(3)}\!R_\varphi^{\,\varphi}\!\simeq\! \frac{\Delta}{a^2r^2}
,\;
\Delta \!=\! \frac{8\pi G_N\phi_0^2}{c^4}
,
\label{Ricci scalar: asymptotic}
\end{equation}
\vskip -0.2cm\noindent where $\Delta$ denotes the solid deficit
angle.

\vskip -1.cm
\section*{Acknowledgement}
\vskip -0.3cm
This work is part of the D-ITP consortium, a program of the
Netherlands Organization for Scientific Research (NWO) that is
funded by the Dutch Ministry of Education, Culture and Science
(OCW). A.M. is funded by NEWFELPRO,  an International Fellowship
Mobility Programme for Experienced Researchers in Croatia and by the
D-ITP.
 The authors thank Jose Maciel Natario for a very useful
comment. 

\end{document}